\documentclass[aps, prd, reprint, twocolumn, superscriptaddress, nofootinbib, nobibnotes, floatfix, nobalancelastpage]{revtex4-2}

\usepackage{amsmath}
\usepackage{amssymb}
\usepackage{graphicx}
\usepackage{xcolor}
\usepackage{footnote}

\begin{document}

\title{Ab Initio Cosmological Simulations:\\ From Inflation to Present-Day Structure Formation}

\author{Drew Jamieson}
\email{jamieson@mpa-garching.mpg.de}
\affiliation{Max-Planck-Institut f\"ur Astrophysik, Karl-Schwarzschild-Straße 1, 85748 Garching, Germany}

\author{Angelo Caravano}
\email{a.caravano@uva.nl}
\affiliation{Gravitation and Astroparticle Physics (GRAPPA), University of Amsterdam
Science Park 904, 1098 XH Amsterdam, Netherlands}

\author{Eiichiro Komatsu}
\affiliation{Max-Planck-Institut f\"ur Astrophysik, Karl-Schwarzschild-Straße 1, 85748 Garching, Germany}
\affiliation{
Ludwig-Maximilians-Universit\"at M\"unchen, Schellingstr. 4, 80799 M\"unchen, Germany
}
\affiliation{Kavli Institute for the Physics and Mathematics of the Universe (Kavli IPMU, WPI), Todai Institutes for Advanced Study, The University of Tokyo, Kashiwa 277-8583, Japan}

\date{\today}

\begin{abstract}
The cosmic web preserves a record of the physics that shaped the universe in its earliest moments, the period of exponential expansion known as cosmic inflation. However, if inflation involves significant nonlinear interactions, there are no direct theoretical predictions for the resulting cosmic web. We present the first simulation of the entire history of the universe, from deep in the inflationary epoch to the present-day cosmic structure. Applying this to axion-U(1) inflation, we find that early-universe interactions enhance small-scale structure at high redshift, imprinting the matter power spectrum and the mass function of dark matter halos with signatures of a modified primordial curvature power spectrum and non-Gaussianity. These signatures make high-redshift galaxy surveys, 21-cm observations, and line-intensity mapping promising probes of inflationary physics. More broadly, this \textit{ab initio} approach provides a novel framework for mapping the signatures of nonlinear inflationary dynamics onto observable cosmic structures across cosmic time.
\end{abstract}

\maketitle

\section{\label{sec:intro}Introduction}

The initial condition for the large-scale distribution of galaxies was generated in the early universe by physics that is largely unknown~\cite{Baumann:2022mni}. Its specific degrees of freedom and interactions left their mark as subtle but measurable statistical properties in the galaxy field. To test theoretical models of the early universe, we must accurately determine what those models predict about the origin of cosmic structure and connect those predictions to observational survey data.

Studies of the cosmic microwave background (CMB) and galaxy distribution were instrumental in establishing a standard model of cosmology~\cite{WMAP:2003elm,SDSS:2003eyi,Planck:2018vyg}. In the standard picture, a brief period of near-exponential expansion known as \textit{cosmic inflation} \cite{Starobinsky:1980te,Guth:1980zm,Sato:1981qmu,Albrecht:1982wi,Linde:1981mu} stretched quantum vacuum fluctuations to cosmic scales, transforming them into density perturbations that form the earliest cosmic structures \cite{mukhanov/chibisov:1981,hawking:1982,starobinsky:1982,guth/pi:1982,bardeen/turner/steinhardt:1983}. The near-scale-invariant primordial curvature power spectrum and high level of Gaussianity observed in the CMB support this general picture~\cite{Komatsu:2014ioa}. Going beyond these results and identifying specific details of early-universe physics requires detecting the nonlinear processes that imprint density perturbations with primordial non-Gaussianity (PNG)~\cite{Bartolo:2004if}.

Searches for PNG in cosmological survey data typically compare observations with a limited set of semi-analytic, perturbatively computed primordial correlation templates, which capture only part of the information encoded in the full primordial density field~\cite{Maldacena:2002vr,Chen:2010xka,Desjacques:2010jw,Komatsu:2010hc}. Recent developments are pushing cosmological analyses beyond this template-based approach. High-redshift 21-cm surveys are opening a three-dimensional window onto a nonlinear tracer of structure~\cite{Furlanetto:2006jb,HERA:2025ajm}, accessing scales near the limits of perturbative bias descriptions~\cite{Qin:2022xho}. In parallel, new field-level inference techniques can extract information beyond summary statistics~\cite{Andrews:2022nvv,Nguyen:2024yth,Andrews:2026iyz}, but applying them to PNG requires a full model of the primordial field.

In addition, interest is growing for inflationary models that resist standard analytic and semi-analytic treatments. An example is axion-U(1) inflation, in which a pseudoscalar axionlike field acts as the inflaton and couples to a U(1) gauge field through a Chern-Simons interaction~\cite{Anber:2009ua,Maleknejad:2012fw,Pajer:2013fsa}. Abundant gauge-field production in this model can invalidate standard perturbative treatments, and lattice simulations of inflation have emerged as an indispensable tool for studying its dynamics~\cite{Caravano:2022epk,Figueroa:2023oxc,Figueroa:2024rkr,Sharma:2024nfu,Iarygina:2025ncl,Jamieson:2025ngu}.

In this work, we present the first cosmological $N$-body simulations initialized from lattice simulations of inflation. By simulating the nonlinear dynamics of axion-U(1) inflation, we generate primordial fields beyond the reach of standard initial-condition algorithms, characterized by a nontrivial hierarchy of non-Gaussian $N$-point correlations that are neither scale invariant nor separable~\cite{Caravano:2022epk,Jamieson:2025ngu}. Together, the lattice and $N$-body simulations follow the entire history of cosmic structure, delivering direct, field-level predictions from inflation.

\section{\label{sec:sims}\textit{Ab initio} simulations}

Numerical studies of late-time cosmic structure formation typically begin long after inflation ends. Density perturbations are initialized either as a Gaussian random field drawn from the primordial power spectrum, parameterized by the scalar amplitude $A_{\rm s}$ and spectral tilt $n_{\rm s}$, or as a slightly non-Gaussian field with an assumed form of PNG from a standard set of scale-invariant bispectrum templates with the local, equilateral, or orthogonal shapes \cite{Scoccimarro:2011pz,Coulton:2022qbc}. Here, we take a more fundamental approach, simulating the full nonlinear dynamics of inflation to generate these initial conditions.

\begin{figure}
\centering
\includegraphics[width=\linewidth]{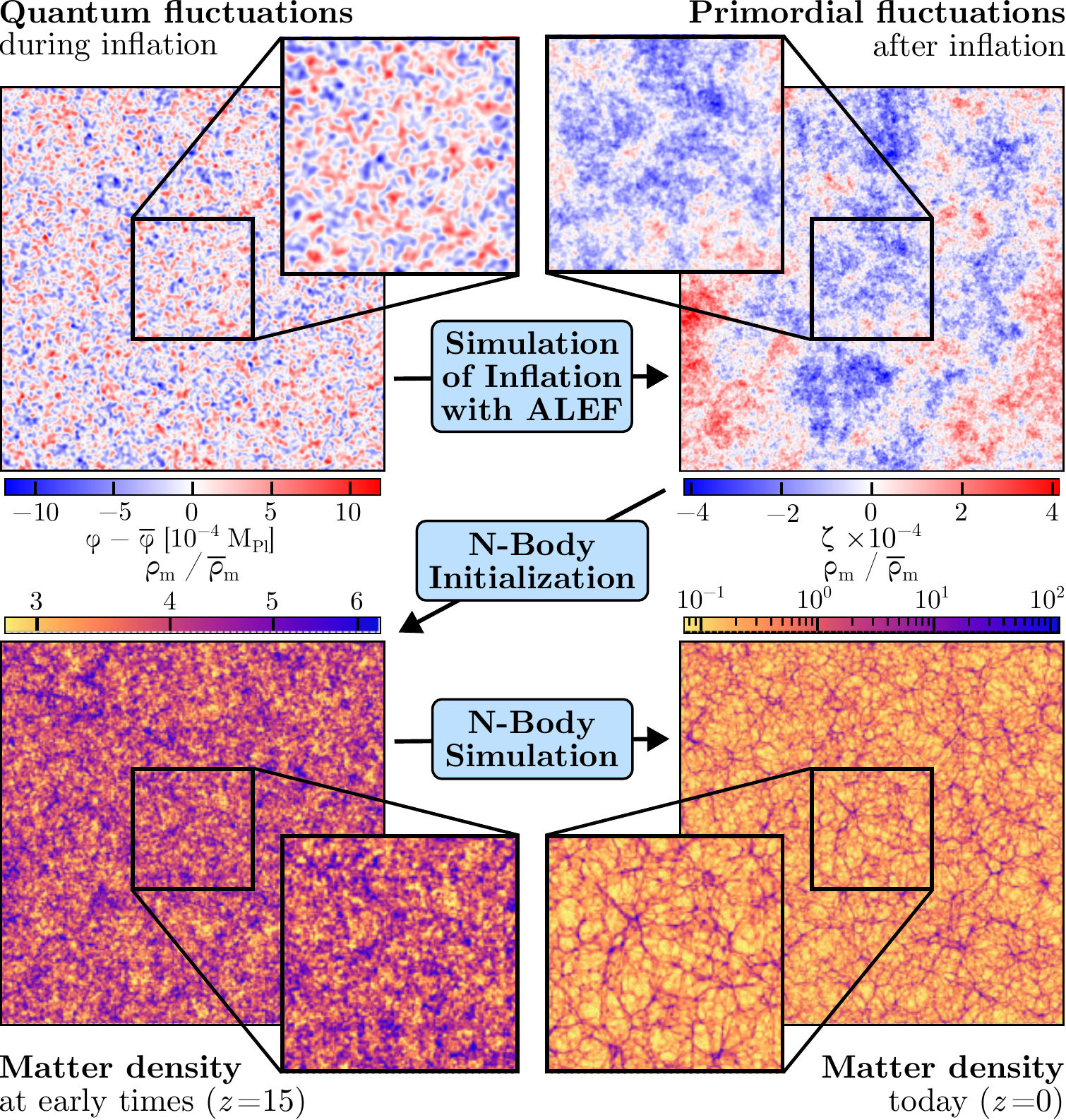}
\caption{Field-level view of our simulation pipeline. The \textsc{ALEF} lattice simulation evolves the quantum fluctuations of the inflaton field $\phi$ deep in the inflationary epoch (top left) into the primordial curvature perturbations $\zeta$ at the end of inflation (top right). These are transformed into a linear matter density field $\rho_{\rm m}$ that initializes the $N$-body simulation, and is evolved from early times (bottom left, $z=15$) to the present-day cosmic web (bottom right, $z=0$).}
\label{fig:slices}
\end{figure}

Adiabatic perturbations generated during inflation are conserved on scales larger than the Hubble horizon size through the end of inflation and reheating~\cite{Weinberg:2003sw}. After reentering the Hubble horizon, they evolve linearly due to their small amplitude, until they begin to cluster gravitationally. We can therefore transform the three-dimensional field of curvature perturbations, $\zeta(\mathbf{x})$, at the end of inflation into linear matter density perturbations by convolving them with the linear matter transfer function at the field level. Using this field as the initial condition for an $N$-body simulation, we follow the entire nonlinear evolution of cosmic structure formation self-consistently from inflation to the present day. Fig.~\ref{fig:slices} shows this pipeline, from inflation to the late-time cosmic web.

We simulate inflation using a newly developed code called Adaptive Lattice Evolved Fields (ALEF) \cite{Jamieson:2025ngu}. The code simulates inflationary field theories on a discrete, uniform grid in comoving coordinates with $N_{\rm grid}^3$ sites and a finite box length $L_{\rm box}$, with periodic boundary conditions. The grid represents an infrared (IR) and ultraviolet (UV) regulated field theory, where the IR cutoff is imposed by the fundamental wave number $k_{\rm F} = 2\pi/L_{\rm box}$, and the UV cutoff is imposed by the Nyquist wave number $k_{\rm Ny} = k_{\rm F} N_{\rm grid}/2$.

During inflation, the inflaton field $\phi$ dominates the energy density of the universe through its potential energy $V(\phi)$, which sets the Hubble rate of expansion, $H$, and determines the comoving Hubble horizon size, $(aH)^{-1}$. We set up our initial conditions when the fundamental mode $k_{F}$ is sufficiently inside the horizon, $k_F\gg aH$. The fields are initialized as Gaussian random fields drawn from the mode functions of the linearized field equations, which represent a sample of the Bunch-Davies vacuum. We begin with a low-resolution grid, $N_{\rm grid}=16$. When $aH$ grows to $k_{\rm Ny}/e^4$, we double the grid resolution and initialize the new high-frequency modes in the same way. We repeat this temporal mesh refinement process until we reach our target resolution of $N_{\rm grid}=512$.

The classical field equations are integrated in Fourier space using an embedded 7(6) Runge-Kutta method with adaptive time stepping. Gradients are evaluated in Fourier space and nonlinear field interactions in real space. We stop simulations when all modes are frozen outside the horizon. Our method of temporal mesh refinement captures the relevant dynamic range of nonlinear mode couplings while avoiding costly numerical simulation of small-scale linear dynamics during early stages of the simulation. We have verified that our results are converged with respect to mesh refinement time, grid resolution, and stop time. Further details of our numerical methods and convergence tests are described in Ref.~\cite{Jamieson:2025ngu}.

\section{\label{sec:au1}Axion-U(1) Inflation}

The model we study is axion-U(1) inflation~\cite{Anber:2009ua}, in which a pseudoscalar axionlike field acts as the inflaton and couples to a U(1) gauge field through a Chern-Simons interaction. Assuming a background Friedmann-Lema\^itre-Robertson-Walker metric, the nonlinear field equations of this system are
\begin{align}
-\square \phi = -2 \mathcal{H} \phi' - a^2 (V_{,\phi} + \Delta m_{\rm eff}^2 \delta \phi) + \frac{g_{\rm CS}}{a^{2}} \vec{E}\cdot\vec{B}, \\
-\square \varphi = g_{\rm CS} \vec{\nabla} \phi \cdot \vec{B}, \\
-\square \vec{A} = -g_{\rm CS} (\phi' \vec{B} + \vec{\nabla} \phi \times \vec{E}).
\end{align}
Here, $\phi$ is the axionlike inflaton field and $A_{\mu} = (-\varphi, \vec{A})$ is the U(1) gauge potential in the Lorenz gauge $\partial^\mu A_{\mu}=0$. The magnetic and electric fields are defined as $\vec{B} = \vec{\nabla} \times \vec{A}$ and $\vec{E} = -\vec{A}' - \vec{\nabla} \varphi$, respectively, and $g_{\rm CS}$ is the Chern-Simons coupling strength. $V_{,\phi}$ is the slope of the inflaton potential and $\delta\phi=\phi- \bar{\phi}$. The wave operator is $\square f = -f'' + \nabla^2 f$, with primes denoting derivatives with respect to conformal time $\tau$. The scale factor $a$ has a conformal Hubble rate $\mathcal{H} = a'/a=aH$ and is evolved in ALEF using the Friedman equations. The effects of scalar metric perturbations are included linearly via the inflaton mass shift, $\Delta m_{\rm eff}^2 = -6 \mathcal{H}'$ (for more details, see Ref.~\cite{Jamieson:2025ngu}).

The phenomenology of this model has been studied using analytical and semi-analytical methods (see, e.g., Refs.~\cite{Barnaby:2010vf,Anber:2012du,Anber:2009ua,Gorbar:2021rlt,Peloso:2022ovc,Barbon:2025wjl}). First, the slowly rolling axionlike inflaton triggers a chiral instability in the linearized gauge field equations, exponentially exciting one of the two U(1) helicity states. The excited gauge field imprints nonlinear fluctuations on the inflaton through the Chern-Simons interaction $g_{\rm CS}\vec{E}\cdot\vec{B}$. These sourced fluctuations represent parity-violating PNG that modify the primordial power spectrum and generate a nonzero bispectrum, trispectrum, and higher $N$-point correlations. Eventually, when the mean of $\vec{E}\cdot\vec{B}$ grows large enough, the system enters a strong backreaction regime in which gauge field fluctuations significantly modify the background inflaton evolution. Here we focus on the generation of PNG in the weak backreaction regime.

\section{\label{sec:nbody}$N$-body Simulations}

At the end of our inflation simulations, the inflaton fluctuations are small enough ($|\delta \phi / \bar{\phi}|\sim10^{-5}$) that they can be linearly converted to curvature perturbations, $\zeta = -(\mathcal{H}/\bar{\phi}') \delta \phi$,
which are conserved on superhorizon scales. The gauge field fluctuations, being vector perturbations, decay on superhorizon scales and do not play a role in later epochs of structure formation.

The curvature perturbations have both vacuum and sourced contributions $\zeta = \zeta_{\rm vac} + \zeta_{\rm src}$. The vacuum contributions arise from the linear field equations and dominate the power spectrum on large scales, yielding negligible PNG. We isolate the sourced and vacuum contributions by running a pair of \textsc{ALEF} simulations with the same random seed and setting $g_{\rm CS}=0$ for one of them. In this way, we can isolate the impact of the nonlinear axion-gauge interaction on large-scale structure by running $N$-body simulations with initial conditions with and without $\zeta_{\rm src}$. We also run a third set of simulations with Gaussian initial conditions matched to the primordial power spectrum of the full axion-U(1) simulations. We construct them by linearly rescaling the vacuum initial conditions from the $g_{\rm CS}=0$ \textsc{ALEF} simulation. We label the three sets as ``G'' (Gaussian), ``SF'' (single field, $\zeta=\zeta_{\rm vac}$), and ``AU1'' (axion-U(1), $\zeta=\zeta_{\rm vac}+\zeta_{\rm src}$). A field-level view of the differences between these three sets of simulations is provided by density slices in Appendix~\ref{supp:slices}, and the primordial power spectrum, bispectrum, and trispectrum measured from these fields are presented in Appendix~\ref{supp:stats}.

We transform the curvature perturbations to linear matter density perturbations in Fourier space, $\delta_{\rm m}(\tau, \vec{k}) = D(\tau) \zeta(\tau, \vec{k}) T_{\rm m}(\vec{k})$, where $T_{\rm m}(\vec{k})$ is the matter transfer function, which we compute numerically at redshift $z=0$ using the cosmological Boltzmann solver CLASS \cite{Blas:2011rf}. $D(\tau)$ is the linear growth factor normalized to unity at $z=0$.

Next, we initialize the $N$-body simulation through Lagrangian perturbation theory. Each grid site of our linear density field\footnote{Here, by \textit{linear}, we mean linear with respect to late-time gravitational evolution. The SF and AU1 fields have nonlinear contributions from the inflation simulations.} is treated as a particle with initial grid position $\vec{q}$ and initial displacement $\vec{\Psi}(\vec{q})$. Linear order displacements are given by the Zeldovich approximation, $\vec{\Psi} = -D(\tau) \nabla^{-2} \nabla \delta_{\rm m}(\tau, \vec{q})$. Higher-order corrections then follow from Lagrangian perturbation theory. We use the public code \textsc{Music2-MonofonIC} \cite{Michaux:2020yis}, modified to accept our linear $\delta_{\rm m}$ field as input, to compute the third-order (3LPT) initial particle displacements and velocities at $z=15$. We then use the public $N$-body code \textsc{gadget-4} \cite{Springel:2020plp} to evolve the $N$-body system until $z=0$.

We ran simulations at two box lengths: $L_{\rm box} = 100~{\rm Mpc}$ with high-redshift snapshots at $z=$ 12, 10, 8, and 6, and $L_{\rm box} = 1000~{\rm Mpc}$ with low-redshift snapshots at $z=$ 3, 2, 1, 0.5, and 0. Each simulation has $4\times512^3$ particles, initialized from an $N_{\rm grid}=512^3$ \textsc{ALEF} simulation using the \textsc{Music2-MonofonIC} face-centered-cubic grid scheme. For each box length, we run three sets of ten simulations with the same random seed but different initial conditions: AU1, SF, and G. The simulation parameters are summarized in Table~\ref{tab:sims}.

\begin{table}[htbp]
\centering
\begin{tabular}{lcc}
\hline\hline
Parameter & Small Box & Large Box \\
\hline
$L_{\rm box}$ & $100~{\rm Mpc}$ & $1000~{\rm Mpc}$ \\
$N_{\rm part}$ & $4\times512^3$ & $4\times512^3$ \\
$N_{\rm grid}$ & $512^3$ & $512^3$ \\
IC type & AU1, SF, G & AU1, SF, G \\
Realizations & 10 & 10 \\
Snapshots ($z$) & (12, 10, 8, 6) & (3, 2, 1, 0.5, 0) \\
\hline
\end{tabular}
\caption{Summary of $N$-body simulation parameters. For each box length, three sets of ten simulations are run with axion-U(1) (AU1), single field (SF), and Gaussian (G) initial conditions.}
\label{tab:sims}
\end{table}

For the \textsc{ALEF} simulation, we choose an $\alpha$-attractor potential~\cite{Kallosh:2013yoa},
\begin{align}
V(\phi) = \frac{M_{\phi}^2 M^2_{\rm Pl}}{2}\Big(1 - e^{-\alpha_{V} \phi/ M_{\rm Pl}} \Big)^2,
\end{align}
where $M_{\phi}=4.653\times10^{-6}~M_{\rm Pl}$, with $M_{\rm Pl}$ the reduced Planck mass and $\alpha_{V} = \sqrt{20/3}$. This fixes the vacuum power spectrum to match the \textit{Planck} 2018 best-fit parameters \cite{Planck:2018vyg} with a negligible tensor-to-scalar ratio. We fix the Chern-Simons coupling to $g_{\rm CS}=750~M_{\rm Pl}^{-1}$, which saturates the $2\sigma$ upper bound on the bispectrum based on the \textit{Planck} 2018 constraints on the scale-invariant equilateral bispectrum template \cite{Planck:2019kim,Jamieson:2025ngu}. The bispectrum for this mode peaks for equilateral configurations, but has a significant blue tilt due to the Chern-Simons coupling breaking scale invariance, and the overall shape does not have a standard, separable form. By initializing our $N$-body simulations with \textsc{ALEF}-simulated primordial curvature fluctuations, we can study the full impact of PNG sourced by the physics of the axion-gauge coupling.

For the $N$-body simulation, we choose cosmological parameters matching the $\Lambda$ cold dark matter ($\Lambda$CDM) best-fit parameters from \textit{Planck} 2018: $\Omega_{\rm m}=0.315$, $\Omega_{\rm b}=0.049$, and $h=0.6774$~\cite{Planck:2018vyg}. The axion-U(1) model that we simulate with axion-gauge coupling strength $g_{\rm CS}$ represents a one-parameter extension to the standard cosmological model.

\section{\label{sec:res}Results}

\begin{figure*}
\centering
\includegraphics[width=0.45\linewidth]{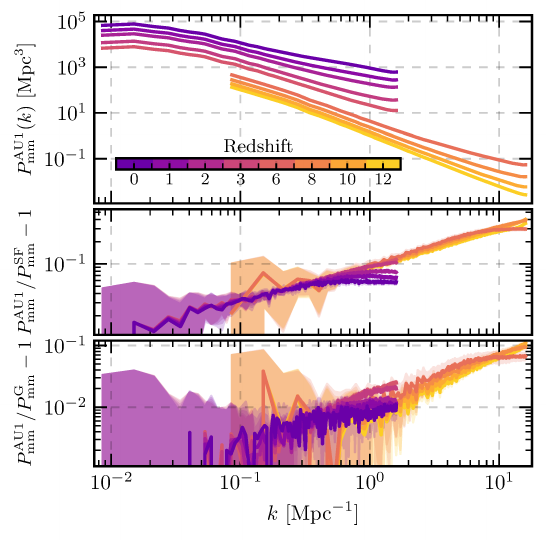}
\hfill
\includegraphics[width=0.45\linewidth]{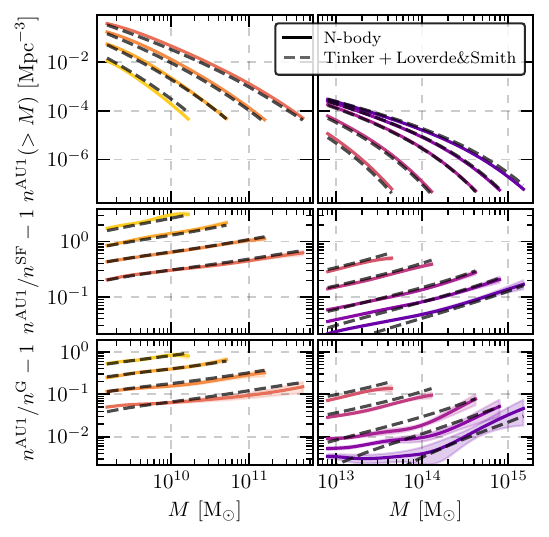}
\hfill
\hfill
\caption{Matter power spectra (left) and halo mass functions (right) from $N$-body simulations with initial conditions from lattice simulations of inflation, for the high-redshift small-box simulations and the low-redshift large-box simulations. Within each column, the upper panel shows results from axion-U(1) (AU1) inflation initial conditions, the middle panel shows the fractional difference between AU1 and single-field (SF) inflation, and the lower panel shows the fractional difference between AU1 and Gaussian (G) initial conditions, where the initial Gaussian power spectrum matches the AU1 primordial power spectrum. Dashed lines in the right panels show the prediction from the Loverde-Smith log-Edgeworth expansion \cite{LoVerde:2011iz} using the Gaussian mass function from Tinker \textit{et al.}~\cite{Tinker:2008ff}.}
\label{fig:results}
\end{figure*}

\subsection{\label{subsec:respk} Matter Power Spectrum}

The nonlinear matter power spectra are displayed in the left panel of Fig.~\ref{fig:results}. Comparison with the SF simulations indicates the impact of the axion-gauge coupling on structure formation, while comparison with the G simulations isolates the impact of PNG.

AU1 simulations exhibit enhanced small-scale power relative to both SF and G. This reflects not only the additional primordial power from $\zeta_{\rm src}$ but also a further nonlinear enhancement from its PNG.

Compared with the SF power spectrum, the AU1 power enhancement is about 10\% at $k=1~h/{\rm Mpc}$ for redshifts $z\ge2$, and 40\% at $k=10~h/{\rm Mpc}$ for redshifts $z\ge8$. The effect due to PNG is 2\% at $k=1~h/{\rm Mpc}$ for redshifts $z\ge2$ and 10\% at $k=10~h/{\rm Mpc}$ for redshifts $z\ge8$. Both the linear and PNG effects are damped by nonlinear gravitational clustering, as seen in the lowest-redshift power spectrum on the smallest measured scales in Fig.~\ref{fig:results}. This nonlinear damping demonstrates that the tightest constraints on axion-U(1) inflation from the matter power spectrum can be obtained from high-redshift observations of the large-scale structure on quasilinear scales.

\subsection{\label{subsec:hmf} Halo Mass Function}

Dark matter halos form from the collapse of overdense peaks in the primordial density field. Galaxies form within dark matter halos, so the abundance and distribution of halos are closely related to the observed distribution of galaxies~\cite{Mo:2010}. The halo mass function quantifies the abundance of halos above a given mass threshold $M$ and is a sensitive probe of the statistics of the primordial density field~\cite{Lucchin:1987yv,Matarrese:2000iz}.

Before presenting the results of our $N$-body simulations on the halo mass function, we provide a physical explanation of how it can be sensitive to different primordial initial conditions.
The Press-Schechter formalism connects the statistical distribution of peaks in the linear density field to the late-time abundance of halos \cite{Press:1973iz}. In this formalism, density peaks are treated as isolated, spherical overdensities that collapse to form virialized halos. The halo mass function is given by
\begin{align}
n(>M) = \bar{\rho}_{\rm m} \int_M^{\infty} \frac{{\rm d}M'}{M'} f(M') \, ,
\end{align}
where $\bar{\rho}_{\rm m}$ is the mean matter density and $f(M)$ is the differential fraction of mass in halos of mass $M$. The original Press-Schechter formalism assumes a Gaussian linear density field, but it has been extended to account for PNG effects. The calculation of Loverde \& Smith \cite{LoVerde:2011iz} showed $f(M) = f_{\rm G}(M) R(M)$, where $f_{\rm G}$ is the Gaussian mass fraction and $R$ is a non-Gaussian correction factor. The Gaussian mass fraction is assumed to depend only on $\sigma^2(M)$, the variance of the linear density field smoothed on a length scale $r = (3 M / (4\pi \bar{\rho}_{\rm m}))^{1/3}$ encompassing mass $M$, which is sensitive to the primordial power spectrum. Typically, $f_{\rm G}$ is calibrated using $N$-body simulations, and here we use the mass function from Tinker \textit{et al.} \cite{Tinker:2008ff}. The non-Gaussian correction factor $R$ depends on the reduced cumulants of the linear density field smoothed on scale $M$, which are related to its connected $N$-point correlation functions, specifically the skewness from the bispectrum and kurtosis from the trispectrum.

The halo mass functions from our simulations are displayed in the right panel of Fig.~\ref{fig:results}. The total effects of the axion-gauge interaction and those due to PNG alone are significantly stronger compared to the effects on the power spectrum. We find elevated halo counts by more than 200\% at $z=12$ compared with single-field inflation and 60\% compared with Gaussian simulations with matched AU1 primordial power. For lower-redshift, large-volume snapshots, we find $\sim10$\% enhancements in halo counts compared with single-field inflation, of which $\sim1$\% is due to PNG alone. The enhancement in the halo mass function has a consistent positive slope, so that the highest mass halos are most sensitive to the axion-gauge interaction and PNG.

The dashed lines in Fig.~\ref{fig:results} show the prediction from the Loverde-Smith log-Edgeworth expansion \cite{LoVerde:2011iz} using the Gaussian mass function from Tinker \textit{et al.} \cite{Tinker:2008ff}. We measured the variance and reduced cumulants in the linear density fields (see Appendix~\ref{supp:smoothed}), averaging over the ten realizations. We find that the log-Edgeworth expansion is in good agreement with our $N$-body simulation results, indicating that the impact of PNG on the halo mass function is consistent with the primordial statistics generated by the nonlinear axion-gauge interaction from our lattice simulations of inflation.

\section{\label{sec:conclusions}Conclusions}

We have run the first cosmological $N$-body simulations initialized with primordial curvature perturbations generated by lattice simulations of inflation. For the inflation simulations, we used a new lattice code, ALEF \cite{Jamieson:2025ngu}, to simulate an axion-U(1) model. These simulations generate realizations of the primordial curvature field imbued with nontrivial statistics that cannot be reproduced using standard initialization algorithms in the literature.

We find enhanced nonlinear matter power on small scales and a substantially stronger enhancement of the halo mass function. The stronger halo response reflects the sensitivity of rare peaks to the higher-order statistics of the primordial field. At $z=12$, halo counts are elevated by more than 200\% relative to SF and remain 60\% higher after matching the primordial power spectrum in Gaussian simulations. This comparison demonstrates that PNG itself contributes substantially to both enhancements. These effects are largest at early times, making high-redshift observations particularly sensitive to nonlinear axion-U(1) dynamics during inflation.

The halo enhancement increases with mass, producing the largest departures for the most massive halos. This result may be relevant to interpreting JWST observations of candidate massive galaxies at high redshift~\cite{2023Natur.616..266L,Boylan-Kolchin:2022kae,Blamart:2025szc} and the abundant compact sources known as ``little red dots''~\cite{Matthee:2023utn,2024ApJ...964...39G}. Connecting these populations to the predicted halo abundances requires modeling galaxy formation beyond our dark-matter-only simulations. These findings identify high-redshift galaxy surveys, 21-cm observations, and other line-intensity mapping experiments as promising complementary probes of the enhanced early structure formation predicted here.

This work establishes a new method for studying the effects of inflationary phenomenology at the field level by simulating its nonlinear dynamics and using these simulations to initialize cosmological $N$-body simulations of large-scale structure formation. Our approach opens pathways for field-level and simulation-based inference schemes to test inflationary models, allowing us to determine what inflation really predicts about the origin of cosmic structure.

\section*{Acknowledgments}
This project has received funding from the European Union's Horizon Europe research and innovation programme under Marie Skłodowska-Curie grant agreement No. 101202657. This work was also supported in part by the Deutsche Forschungsgemeinschaft (DFG, German Research Foundation) under Germany's Excellence Strategy - EXC-2094/2 - 390783311 and the European Research Council (ERC) under the European Union's Horizon 2020 research and innovation program (NewPhysCMB, Grant agreement No. 101264428). Views and opinions expressed are however those of the authors only and do not necessarily reflect those of the European Union or the ERC Executive Agency. Neither the European Union nor the granting authority can be held responsible for them.

\appendix

\section{\label{supp:slices}Density Field Slices}

For each simulation snapshot, we construct an estimator of the 3D density field by distributing the $4\times512^3$ $N$-body particles onto a uniform grid. To plot slices of the density field, we select a slab of thickness $L_{\rm box} / 32$, around the center of the simulation box in the $z$-direction and $L_{\rm box}/4$ in the $x$ and $y$ directions. We construct a $k$-d tree, finding the 16th nearest neighbor of each particle in the slab, and use the distance to this neighbor to define an adaptive smoothing length. Each particle is then distributed onto a grid of size $4096\times4096\times128$ with a Gaussian smoothing kernel the size of its smoothing length. We average over the 16 voxels around the center of the slab in the $z$-direction to obtain an average 2D density slice. We also zoom into a patch of length $L_{\rm box}/16$, keeping the grid size fixed for higher resolution.

\begin{figure*}
\includegraphics[width=0.49\linewidth]{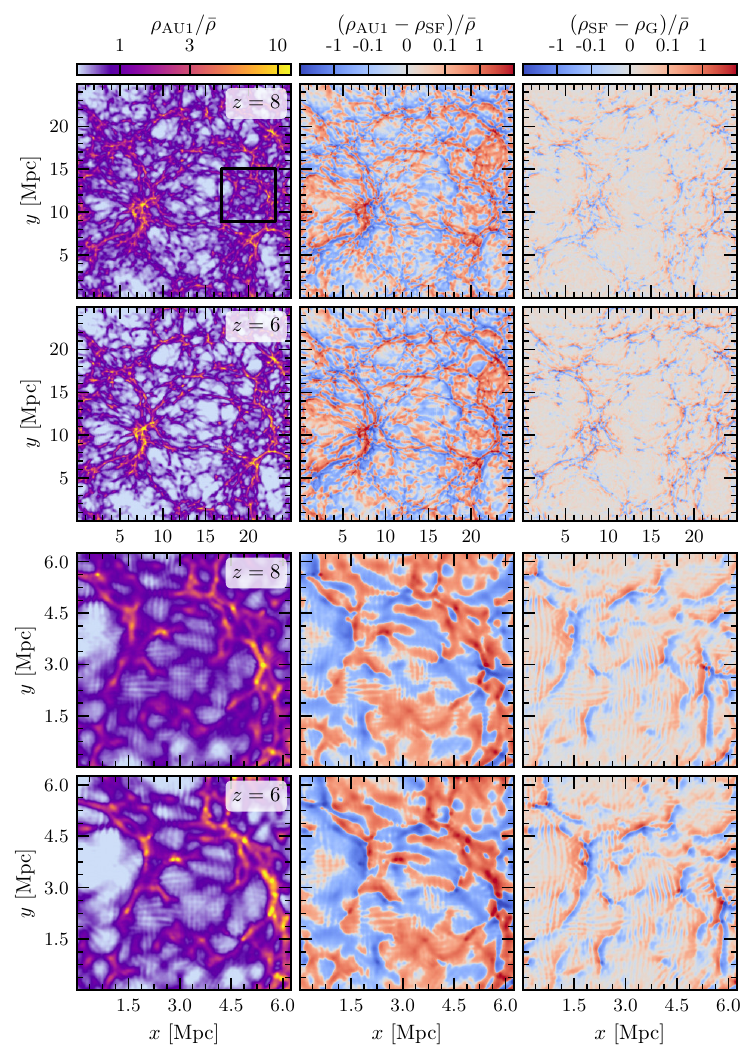}
\includegraphics[width=0.49\linewidth]{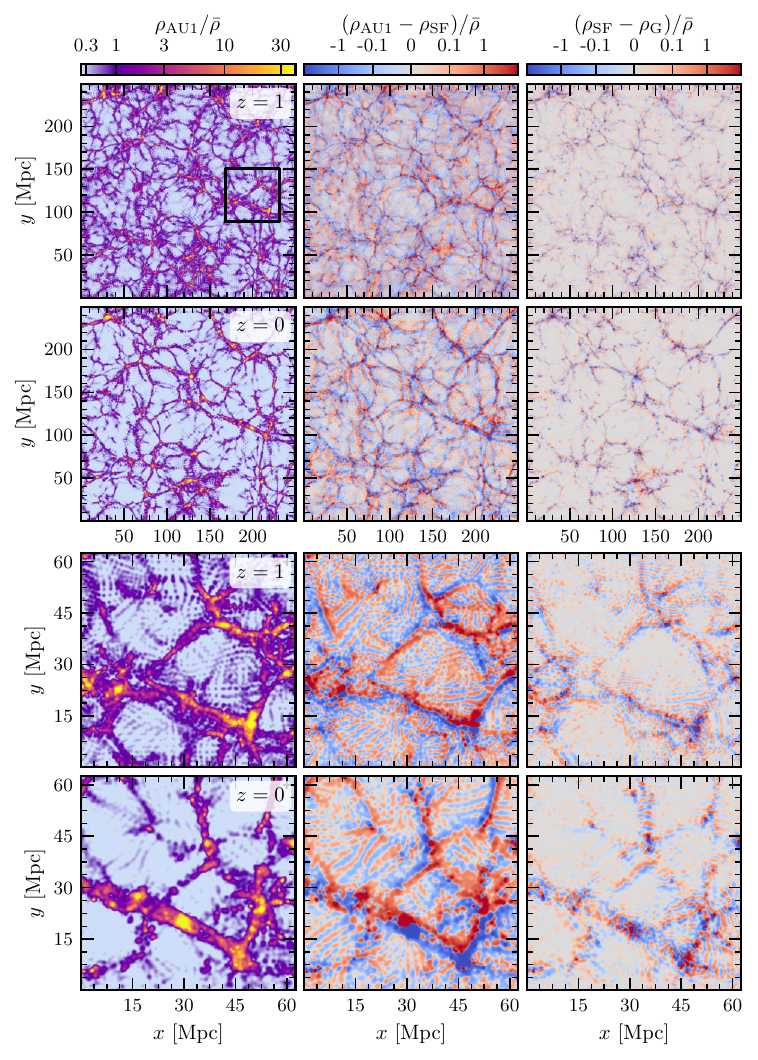}
\caption{Slices of the nonlinear density field from $N$-body simulations with box lengths $L_{\rm box}=100\ {\rm Mpc}$ (left) and $L_{\rm box}=1000\ {\rm Mpc}$ (right). In each of the four sets of panels, the leftmost panels show the density field from the AU1 simulations, the middle panels show the difference in density between the AU1 and SF simulations, and the rightmost panels show the difference in density between the SF and G simulations. The top two sets of panels show a patch of width $L_{\rm box}/4$ in the $x$ and $y$ directions, with a black square in the top-left panel marking the zoomed-in region shown in the bottom two sets of panels, which have width $L_{\rm box}/16$. The left sets of panels show high redshifts $z=8$ and $z=6$ from an $L_{\rm box} = 100~{\rm Mpc}$ simulation and the right sets show the low redshifts $z=1$ and $z=0$ from an $L_{\rm box} = 1000~{\rm Mpc}$ simulation.}
\label{fig:supp:slices}
\end{figure*}

The 2D slices through the density field are displayed in Fig.~\ref{fig:supp:slices} at two redshifts for each box length. In each of the four sets of panels, the leftmost panels show the density field relative to the mean background density for the Axion-U(1) (AU1) simulations. The middle columns show the difference in density between the AU1 simulation and its single field (SF) equivalent simulation. The rightmost panels show the difference in density between the SF and Gaussian (G) density fields. The differences between the AU1 and SF densities are spatially distributed more evenly through the box at high redshift. The differences become more concentrated around high-density, collapsed regions at late times. The late-time differences are also characterized by slight phase differences, or offsets in where high-density features of filaments and halos are located, rather than the overall density of these objects. The differences between the SF and G density field are smaller, are always concentrated around collapsed regions, and are always characterized by offsets in the positions of the collapsed high-density objects. This demonstrates that, at the level of the density field, the AU1, SF, and G simulations have large-scale structure that agrees on large scales and deviates only on small scales due to the scale-dependent, blue-tilted statistics of the AU1 initial conditions.

\section{\label{supp:stats}Primordial Statistics}

Our ALEF AU1 lattice inflation simulations generate 3D primordial curvature fluctuation fields with contributions from both linear, Gaussian vacuum fluctuations $\zeta_{\rm vac}$, and nonlinear, non-Gaussian sourced fluctuations $\zeta_{\rm src}$. The total curvature fluctuation is $\zeta = \zeta_{\rm vac} + \zeta_{\rm src}$. By running a pair of simulations with the same initial seed but where one is a single field simulation with the axion-gauge coupling strength $g_{\rm CS}$ set to zero, we isolate the sourced and vacuum fluctuations:
\begin{align}
    \zeta_{\rm vac} = \zeta_{\rm SF} \, , \\
    \zeta_{\rm src} = \zeta_{\rm AU1} - \zeta_{\rm SF} \, .
\end{align}

We measure the primordial $N$-point statistics in $\zeta_{\rm vac}$ and $\zeta_{\rm src}$. The vacuum fluctuations have effectively vanishing non-Gaussianity, so their statistics are characterized entirely by their power spectrum. The cross-correlation between $\zeta_{\rm vac}$ and $\zeta_{\rm src}$ is negligibly small because $\zeta_{\rm src}$ originates from the vacuum fluctuations of the $U(1)$ gauge field, which are uncorrelated with $\zeta_{\rm vac}$. This decomposition into uncorrelated $\zeta_{\rm vac}$ and $\zeta_{\rm src}$ works well because we are far from the strong backreaction regime where nonperturbative mixing between the inflaton and the U(1) gauge field occurs.

\begin{figure*}
\includegraphics[width=0.49\linewidth]{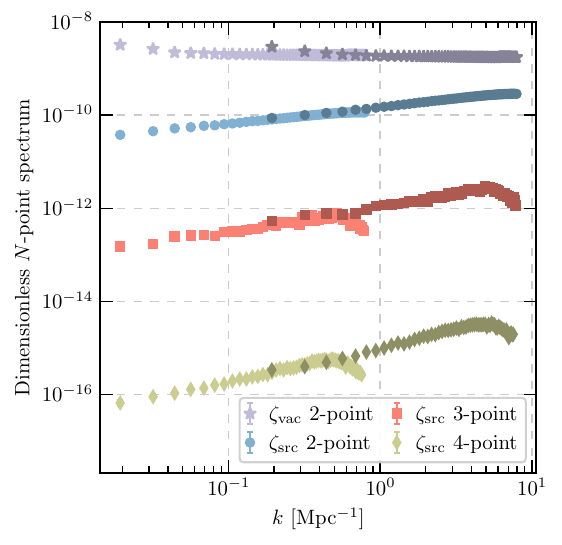}
\includegraphics[width=0.49\linewidth]{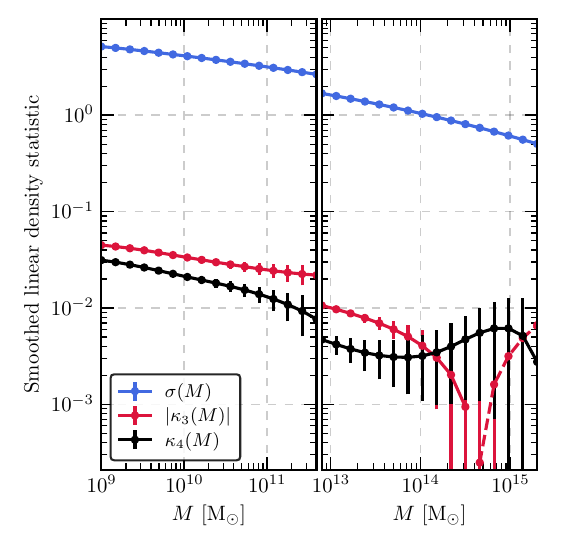}
\caption{Left: $N$-point statistics of the primordial curvature fluctuations from the ALEF AU1 lattice inflation simulations. The lighter large-scale points are measured from the $L_{\rm box} = 1000~{\rm Mpc}$ simulation and the darker small-scale points are measured from the $L_{\rm box} = 100~{\rm Mpc}$ simulations. The bispectrum is measured for equilateral configurations, and the trispectrum is the angle-averaged parity-even component measured for near-regular configurations described in the text. Right: Square root of the variance and reduced cumulants of the smoothed linear matter field initialized from ALEF AU1 simulations. The square-root variance is scaled to redshift $z=0$, while the reduced cumulants are independent of redshift. The dashed lines indicate where the mean values of $\kappa_3(M)$ are negative. Error bars in both panels show the standard error on the mean estimated from the ten realizations.}
\label{fig:supp:npoint}
\end{figure*}

As an illustration of the amplitudes and scale dependence of the AU1 primordial statistics, we plot the power spectrum, bispectrum, and trispectrum in Fig.~\ref{fig:supp:npoint} for certain configurations. The power spectrum, bispectrum, and trispectrum of the curvature fluctuations are defined as 
\begin{widetext}
\begin{align}
\langle \zeta(\vec{k}_1) \zeta(\vec{k}_2) \rangle = (2\pi)^3 \delta^{(3)}_{\rm D}(\vec{k}_1 + \vec{k}_2) P(k_1) \, , \\
\langle \zeta(\vec{k}_1) \zeta(\vec{k}_2) \zeta(\vec{k}_3)  \rangle = (2\pi)^3 \delta^{(3)}_{\rm D}(\vec{k}_1 + \vec{k}_2 +\vec{k}_3) B(k_1, k_2, k_3) \, ,  \\
\langle \zeta(\vec{k}_1) \zeta(\vec{k}_2) \zeta(\vec{k}_3) \zeta(\vec{k}_4) \rangle_{\rm c} = (2\pi)^3 \delta^{(3)}_{\rm D}(\vec{k}_1 + \vec{k}_2 + \vec{k}_3 + \vec{k}_4) T(k_1, k_2, k_3, k_4, \cos(\theta_{12}), \cos(\theta_{23})).
\end{align}
\end{widetext}
In the last equation, the subscript ${\rm c}$ denotes the connected four-point function, which vanishes for a Gaussian field and is obtained by subtracting the three pairwise products of two-point functions from the full four-point function.
The delta functions enforce statistical translational invariance, and the arguments of the $N$-point spectra are (up to reparameterization) determined by isotropy. For the trispectrum, only the connected correlation is taken. We have chosen angular dependence parameterized by $\vec{k}_i \cdot \vec{k}_j = k_i k_j \cos(\theta_{ij})$. There is an additional discrete parameter for the trispectrum given by the sign of the triple vector product ${\rm sign}\big({(\vec{k}_1 \times \vec{k}_2) \cdot \vec{k}_3}\big)$, and we refer to the term in the trispectrum proportional to this sign as its parity-odd component, whereas the term that is independent of this sign is its parity-even component. In general, we expect the chiral instability of the Axion-U(1) model to generate both parity-even and parity-odd components in the trispectrum and higher $N$-point statistics. However, in practice, we have found the parity-odd component to be negligibly small for the angle-averaged statistic we measure here and the parameter choices of our simulations. We defer its study to future work.

We measure these $N$-point spectra by first selecting modes in a spherical Fourier-space shell centered at $k$ with width $\Delta k = 2k_{\rm F}$, where $k_{\rm F}=2\pi/L_{\rm box}$ is the fundamental wavenumber:
\begin{align}
    \zeta_{\rm shell}(\vec{q};k) = \zeta(\vec{q}) \Theta(\vec{q};k)\, ,
\end{align}
where $\Theta(\vec{q};k)$ equals 1 for modes with $\vec{q}$ inside the shell and vanishes otherwise. The power spectrum is estimated as
\begin{align}
P(k) = \frac{1}{N_{P}(k) V_{\rm box}} \sum_{\vec{q}} |\zeta_{\rm shell}(\vec{q};k)|^2 \, ,
\end{align}
and the dimensionless power spectrum is defined as $\Delta(k) = (2 \pi^2)^{-1} k^{3} P(k)$. The normalization counts the number of modes in the shell,
\begin{align}
    N_{P}(k) = \sum_{\vec{q}} \Theta(\vec{q};k) \, .
\end{align}

For the bispectrum and trispectrum, we inverse Fourier transform the shell and the mode selection window to obtain their configuration space fields $\zeta_{\rm shell}(\vec{x};k)$ and $\Theta(\vec{x};k)$, distinguished here from their modes by the argument $\vec{x}$ instead of $\vec{q}$. We estimate the bispectrum for equilateral triangle configurations ($k_1 = k_2 = k_3 = k$),
\begin{align}
    B(k,k,k) = \frac{1}{N_{B}(k,k,k) V_{\rm box}} \sum_{\vec{x}} \zeta_{\rm shell}(\vec{x};k)^3 \, ,
\end{align}
with normalization
\begin{align}
    N_{B}(k,k,k) = \sum_{\vec{x}} \Theta(\vec{x};k)^3 \, .
\end{align}
The dimensionless bispectrum for equilateral configurations is then given by $k^6 B(k,k,k)$.

Finally, we compute the following angle-averaged parity-even trispectrum,
\begin{align}
    T(k_1,k_2,k_3,k_4) &= \frac{1}{N_{T}(k_1,k_2,k_3,k_4) V_{\rm box}} \nonumber \\
    &\quad \times \sum_{\vec{x}} \prod_{n=1}^{4} \zeta_{\rm shell}(\vec{x};k_n) \, ,
\end{align}
with normalization
\begin{align}
    N_{T}(k_1,k_2,k_3,k_4) = \sum_{\vec{x}}  \prod_{n=1}^{4}  \Theta(\vec{x};k_n) \, .
\end{align}
This trispectrum estimator averages over $\cos(\theta_{12})$ and $\cos(\theta_{23})$ for tetrahedra with four side lengths in adjacent, nonoverlapping bins $k_n = k_1 + (n-1) \Delta k$. For these configurations, the purely Gaussian disconnected part of the correlator vanishes due to translation invariance. The dimensionless angle-averaged trispectrum is $k_1^9 T(k_1,k_2,k_3,k_4)$. Since all $k_n$ are determined by $k_1$, we define $k=k_1$ and write this simply as $k^9 T(k)$.

In the left panel of Fig.~\ref{fig:supp:npoint}, we plot the vacuum power spectrum, the sourced power spectrum, the bispectrum, and the trispectrum from our ALEF AU1 simulations. Darker shaded points are from $L_{\rm box}=100\ {\rm Mpc}$ simulations and lighter points are from $L_{\rm box}=1000\ {\rm Mpc}$ simulations. All sourced statistics have a blue tilt, with the slope increasing with the order of the statistic. For the scales in our simulations and for our model parameter choices, we find a clear hierarchy with $\Delta_{\rm src}(k) > k^6 B_{\rm src}(k,k,k) > k^9 T_{\rm src}(k)$. The drop-off in the measured sourced statistics at high $k$ is due to small-scale modes missing above the resolution of our lattice inflation simulations, which would source these scales nonlinearly if included (see Ref.~\cite{Jamieson:2025ngu} for more details). The single field and Gaussian simulations have negligibly small bispectra and connected trispectra.

\section{\label{supp:smoothed}Statistics of the Smoothed Linear Matter Field}

The non-Gaussian halo mass function depends on the variance and the reduced cumulants of the smoothed linear density field. For a linear density field $\delta_{\rm m}(\vec{x})$, we define the smoothed field,
\begin{align}
    \delta_{\rm m}(\vec{x};M) = \int {\rm d}^3y W(|\vec{x}-\vec{y}|;M) \delta_{\rm m}(\vec{y}) \, ,
\end{align}
where the tophat smoothing window for mass $M$ is,
\begin{align}
    W(r;M) = \frac{3}{4\pi R_M^3}\Theta\Big(R_M - r\Big) \, ,
\end{align}
where $\Theta(x)$ is the Heaviside function that vanishes when $x<0$ and is 1 otherwise. The length $R_M = \big(3 M / (4\pi\bar{\rho})\big)^{1/3}$ is the radial size that encompasses mass $M$ at mean density. In Fourier space, the modes of the field
\begin{align}
    \delta_{\rm m}(\vec{k}) = \int {\rm d}^3 x e^{-i\vec{k}\cdot\vec{x}} \delta_{\rm m}(\vec{x}) \, ,
\end{align}
are smoothed by multiplying the Fourier transform of the smoothing kernel
\begin{align}
    W(k;M) = \frac{3}{(k R_M)^3} \Big( \sin(k R_M) - kR_M \cos(k R_M) \Big)\, ,
\end{align}
so
\begin{align}
    \delta_{\rm m}(\vec{k};M) = \delta_{\rm m}(\vec{k}) W(k;M) \, .
\end{align}

The variance of the smoothed density field is
\begin{align}
    \sigma^2(M) = \langle \delta_{\rm m}(\vec{x}; M)^2 \rangle \, ,
\end{align}
and the reduced cumulants are defined as
\begin{align}
    \kappa_n(M) = \frac{\langle \delta_{\rm m}(\vec{x}; M)^n \rangle_{\rm c}}{\sigma^n(M)} \, ,
\end{align}
where $\sigma^n(M)\equiv (\sigma^2(M))^{n/2}$ and only the connected part of the expectation value is taken. Derivatives with respect to $M$ of $\sigma^2(M)$ and the cumulants are formed from terms that involve $\delta_{\rm m}(\vec{x})$ smoothed with derivatives of the window function.
\\ \noindent
The right panel of Fig.~\ref{fig:supp:npoint} displays the mass dependence of $\sigma(M)$, $\kappa_3(M)$, and $\kappa_4(M)$. The reduced cumulants are independent of redshift, and $\sigma(M)$ is shown scaled to redshift $z=0$. Dashed lines indicate where the mean value of $\kappa_3(M)$ becomes negative, although the error bars, which are large at high mass due to cosmic variance and finite box volume, make $\kappa_3(M)$ consistent with zero there. At high mass, $\kappa_4(M)$ appears to dominate over $\kappa_3(M)$. However, here too, the error bars make both quantities consistent with zero. These smoothed statistics of the linear density field and their mass derivatives are the ingredients of the non-Gaussian mass function calculation from Ref.~\cite{LoVerde:2011iz}.

\bibliography{arxiv}
\onecolumngrid
\end{document}